# Coexistence of Rashba Effect and Spin-valley Coupling in TiX$_2$ (X= Te, S and Se) based Heterostructures

Amreen Bano* and Dan Thomas Major

*Department of Chemistry and Institute of Nanotechnology and Advanced Materials (BINA), Bar-Ilan University, Ramat Gan 52900, Israel*

(Dated: December 29, 2022)

Spin-orbit coupling (SOC) combined with broken inversion symmetry play key roles in inducing Rashba effect. The combined spontaneous polarization and Rashba effect enable controlling a material's spin degrees of freedom electrically. In this work we investigated the electronic band structure for several combinations of TiX$_2$ monolayers (X= Te, S and Se): TiTe$_2$/TiSe$_2$, TiTe$_2$/TiS$_2$, and TiSe$_2$/TiS$_2$. Based on the observed orbital hybridization between the different monolayers in these hetero-structures (HSs), we conclude that the most significant Rashba splitting occurs in TiSe$_2$/TiS$_2$. Subsequently, we used Fluorine (F) as an adatom over the surface of TiSe$_2$/TiS$_2$ at hollow and top sites of the surface to enhance the Rashba intensity, as the F adatom induces polarization due to difference in charge distribution. Furthermore, by increasing the number of F atoms on the surface, we reinforced the band splitting, i.e., we observe Rashba splitting accompanied by Zeeman splitting at the valence-band edge states. Berry curvatures at K and K' with equal and opposite nature confirms the existence of valley polarization. The computationally observed properties suggest that these HSs are promising candidates for spin-valley Hall effect devices and other spintronic applications.

## I. INTRODUCTION

Modulating properties such as magnetic ordering[1, 2], superconductivity[3–5] and optical performances [6, 7] in two-dimensional materials has gained immense attention worldwide. Monolayer transition metal dichalcogenides (TMDs), which possess weak inter-layer interactions, can be fabricated with mechanical exfoliation of their parent bulk systems [8]. In absence of inversion symmetry, band splitting at $K$ and $K'$ valleys of H-phase materials [9, 10] make them suitable candidates for valleytronics applications [11, 12]. TMDs based heterostructures or hetero-bilayer systems such as *MoS$_2$*, *WTe$_2$*, and *TiSe$_2$* have attracted the attention of both, experimental and theoretical scientists [13–18]. Despite being weaker than covalent and ionic bonds, van der Waals interaction can significantly influence the electronic properties of a material [8, 19, 20]. Complex two-dimensional materials having large spin-orbit coupling (SOC) play important roles in burgeoning field of spintronics. Spin of an electron is a quantum property which can be used to store and process information [21]. Interaction between an electron's spin and its motion in the local nuclear electric field gives rise to SOC, emphasizing the connection between spin-degree of freedom and orbital motion [21, 22]. Rashba spin-splitting is a phenomenon attributed to broken inversion-symmetry (i.e., out-of-plane), which lifts the spin degeneracy of electronic bands. Rashba effect exists in bulk and low dimensional materials, but remolding of spin carriers is a prominent feature in two-dimensional systems owing to strong SOC. For a two-dimensional electron gas, the Rashba Hamiltonian is given by $H_R = \pm \frac{\hbar^2 k^2_\parallel}{2m} + \alpha_R \sigma \cdot (k_\parallel \times z)$, where $\alpha_R$ is the Rashba constant which indicates the strength of SOC and $\sigma$ represents the Pauli matrices vector [23]. Many phenomena related to the Rashba effect and its tunability have been explored for functional materials [24].

An efficient way to induce and tune the Rashba effect is to fabricate heterostructures. Significant differences in electronegativities of the individual layers of a heterostructure result in internal electric field along the perpendicular direction of heterostructure plane [25]. Very recently, the modified band structures of a 1-T phase TMD heterostructure TiTe$_2$/TiSe$_2$ bilayer was been reported [18] despite of the material lacking translational symmetry. Electronegtivity difference in Se (2.55 on Pauling's scale) and Te (2.1 on Pauling's scale) could lead to local dipole moments which induce Rashba spin-splitting into the material. However, the authors of this work [18] did not explore the spin-valley features of this heterostructure (HS). Motivated by the previously reported research, in the present work we investigated the electronic properties of the 1-T phase TiTe$_2$/TiSe$_2$, TiTe$_2$/TiS$_2$, TiSe$_2$/TiS$_2$ HSs. We show that by modelling different combinations of Ti-based monolayers into hetero-bilayer systems, a generalized Rashba effect can be induced. Additionally, this effect can be enhanced significantly by adding an adatom over the surface. Finally we extracted the Berry curvature from the band structure of the TiSe$_2$/TiS$_2$ HS to confirm the existence of band a splitting at the K and K' valleys, which further proves the coexistence of spintronics (Rashba splitting along the in-plane direction) and valleytronics (spin splitting along the out-of-plane direction).

* banoamreen.7@gmail.com

## II. COMPUTATIONAL METHOD

### A. Calculation Parameters

We performed all calculations using density functional theory (DFT) as implemented in Vienna Ab initio Simulation Package (VASP) [26]. All calculations were conducted with the Perdew–Burke–Ernzerhof (PBE) functional within the framework of the generalized gradient approximation (GGA). To transform rapidly oscillating wavefunctions into smooth wavefunctions, the projector augmented wave method [27, 28] was used. All the calculations were carried out with spin-orbit coupling (SOC). A Γ-centered $6\times6\times1$ k-mesh in a 2D Brillouin zone along with 460 eV kinetic energy cutoff was used for structural relaxations and electronic structure calculations. The convergence criteria during geometry optimization for the total energy and forces were $1\times 10^{-6}$ eV/atom and 0.01 eV/Å, respectively. For the geometry optimized structures, the self consistent field convergence criterion was set to $1\times 10^{-5}$ eV/atom. The calculated lattice constants of the 1T phase of $TiTe_2$, $TiSe_2$ and $TiS_2$ are 3.75 Å, 3.57 Å [18] and 3.46 [29], respectively. The HSs $TiTe_2/TiSe_2$, $TiTe_2/TiS_2$, $TiSe_2/TiS_2$ results constitute a lattice mismatch of 4.8% [30], 7.73% and 3.08%, respectively. A supercell of $3\times3\times1$ of each material was considered to build the HS. To avoid interactions between periodic images along the out-of-plane direction, a thick vacuum of height 22 Å was included. To take into account the van der Waals interaction of the HS monolayers, the DFT-D3 [31] approach was employed. For post-processing of data and plotting, the Pyprocar [32] Python library and Matplotlib [33] were used. To calculate the Berry curvature, the maximally localized Wannier function method as implemented in the Wannier90 package [34] was used to construct real-space maximally localized Wannier functions (MLWFs). To this end, we used the ground state structure of the HSs which were obtained from the Quantum Espresso package [35].

### B. Rashba Model

To explain the Rashba effect induced by spin-chirality, the Bychkov-Rashba Hamiltonian [36, 37] can be used.

$$H_R = -\alpha' \frac{\hbar}{4m^2c^2}\sigma(p \times \nabla_\perp V) \quad (1)$$

where the Rashba primary correlation factor is given by $\alpha'$, $c$ is the speed of light, $m$ is the effective mass, $\nabla_\perp$ denotes the gradient operator and the electric potential is given by $V$.

In 2D geometries the Rashba Hamiltonian becomes (using cross and scalar product construction rules and the electric field ($E_z$) expression)

$$H_R = \frac{\alpha'\hbar^2 E_z}{4m^2c^2}(\sigma \times k)\hat{z} = \alpha(\sigma \times k)\hat{z} \quad (2)$$

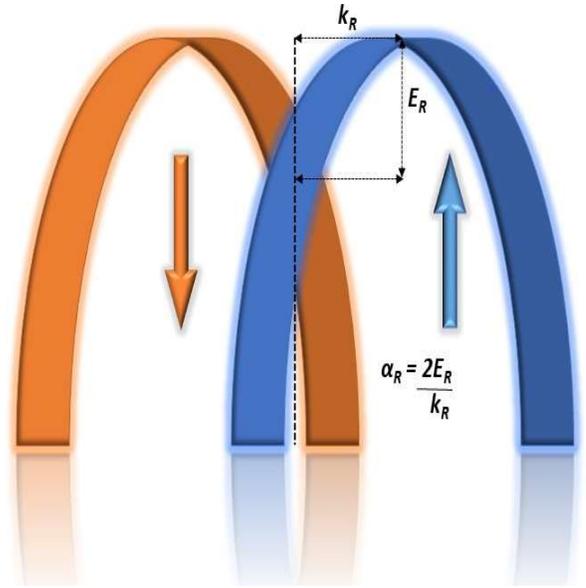

FIG. 1. A diagrammatic representation of the Rashba spin splitting for spin up and spin down chiral states, with Rashba energy $E_R$, momentum offset $k_R$ and Rashba coefficient $\alpha$.

where $E_z$ is the electric field included in the general form of the Rashba interaction coefficient, termed the Rashba constant $\alpha$, which is expressed as

$$\alpha = \frac{\alpha'\hbar^2 E_z}{4m^2c^2} \quad (3)$$

Hence, the dispersion law for the Rashba spin splitting is [37, 38]

$$E_\pm(k) = \frac{\hbar^2 k^2}{2m} \pm \alpha k = \frac{\hbar^2}{2m}(k \pm k_R)^2 - E_R \quad (4)$$

The Rashba energy $E_R$ at momentum offset $k_R$, thus determines the $\alpha$ constant as represented in Fig 1 [39, 40]

$$\alpha = \frac{2E_R}{k_R} \quad (5)$$

## III. RESULTS AND DISCUSSION

### A. Crystal Structure and Stability of the Heterostructures

The various hetero-bilayers constructed from the monolayers of $TiTe_2$, $TiSe_2$ and $TiS_2$ are shown in Fig 2(a-c). The stacking of these HSs plays an important role in achieving the ground state geometry. Lin et al. [18] suggested that AA stacking is energetically preferred over AB and AB' stacking, and we therefore considered AA stacking for all HSs, where all monolayers possess the 1T-phase. The inter-layer spacing, $d$, in $TiTe_2/TiSe_2$, $TiTe_2/TiS_2$, $TiSe_2/TiS_2$ are 6.25 Å [18], 6.04 Å and 5.67

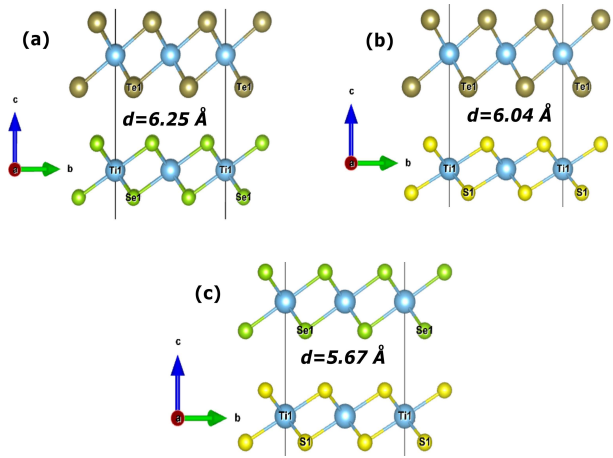

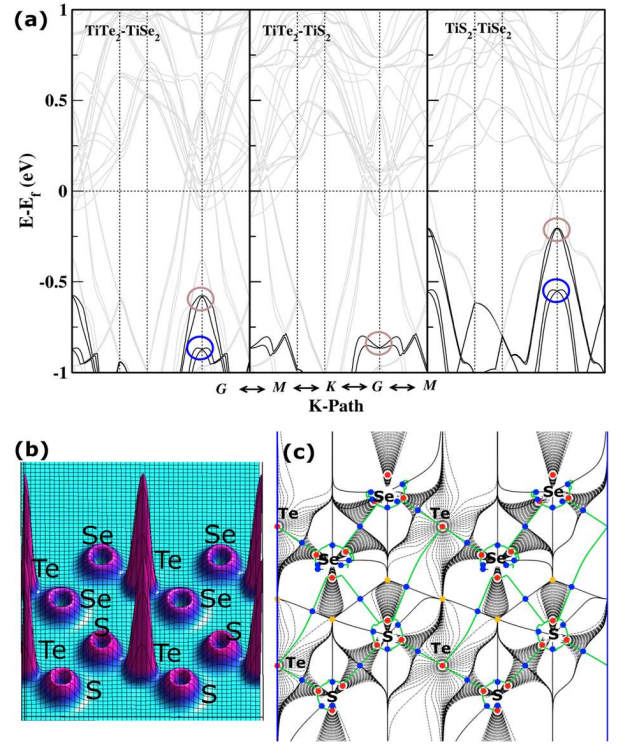

FIG. 2. Heterostructures of a) $TiTe_2/TiSe_2$, b) $TiTe_2/TiS_2$ and c) $TiSe_2/TiS_2$. Vertical spacing $d$ is obtained from energy minimization of each HS.

Å, in good agreement with the reported literature values. In order to examine the structural stability of these HSs, we calculated the binding energy per atom ($E_b$) using the following expression:

$$E_b = \frac{E_{HS} - (E_{ML1} + E_{ML2})}{N} \quad (6)$$

where $E_{HS}$ is the energy of the HS, $E_{ML1}$ and $E_{ML2}$ are the energy of each monolayer forming the HS, and N is the number of atoms in the HS. The binding energy per atom obtained in our calculations are listed in Table 1. We obtained negative binding energies for all three HSs studied in the present work, suggesting that these materials are thermodynamically stable (assuming similar entropy changes upon HS formation).

TABLE I. Binding energy per atom ($E_b$) of calculated and previously reported two-dimensional heterostructures.

| Structure | $E_b$ (eV) |
|---|---|
| $TiTe_2/TiSe_2$ | -0.026 |
| $TiTe_2/TiS_2$ | -0.0441 |
| $TiSe_2/TiS_2$ | -0.0287 |
| $MoSe_2/Zr_2CO_2$ [41] | -0.228 |
| $WSe_2/Zr_2CO_2$ [41] | -0.278 |
| $Gr/HfS_2$ [42] | -0.017 |

## B. Electronic Structure: Origin of Rashba spin-splitting

Electronic band structures of the HSs $TiTe_2/TiSe_2$, $TiTe_2/TiS_2$ and $TiSe_2/TiS_2$, including SOC effects, are shown in Fig 3(a) (from left to right, respectively). Our calculated band structure of $TiTe_2/TiSe_2$ agrees well with the experimentally obtained ARPES maps [18]. To the best of our knowledge, the other HSs have not been characterized experimentally. The bands where Rashba spin splitting around the Γ point can be realised are shown in black in Fig 3(a). From the band splittings we extracted the maximum value of the Rashba parameter obtained using *equation 5* for each HS. The values are 0.287 eV Å (brown circle) and 0.397 (blue circle) in $TiTe_2/TiSe_2$, 0.713 eV Å (brown circle) in $TiTe_2/TiS_2$ and 0.103 eV Å (brown circle) and 0.367 eV Å (blue circle) in $TiSe_2/TiS_2$ HS. Although, these HSs show considerable Rashba band splitting, these are obtained in the deep valance band energy region and are therefore not suitable for practical applications. Moreover, the Rashba splitting obtained for the $TiSe_2/TiS_2$ HS is closer to the Fermi level ($E_F$) than the other HSs. Therefore, our next aim is to tune the $TiSe_2/TiS_2$ HS to achieve Rashba spin-splitting at $E_F$.

FIG. 3. a) Electronic band structure of the heterostructures; $TiTe_2/TiSe_2$ (left), $TiTe_2/TiS_2$ (middle) and $TiSe_2/TiS_2$ (right). Fermi energy is set to 0.0 eV. b) 3D Laplacian charge density map of $TiSe_2/TiS_2$ HS, where magenta color indicates charge accumulation while blue color indicates charge depletion, c) 2D Laplacian charge density contour plot of $TiSe_2/TiS_2$ HS showing maximum (red), minimum (orange) and saddle (blue) points.

In the following, we consider $TiSe_2/TiS_2$ HS for further calculations. Fig 3(b) and 3(c) show the 3D and 2D Laplacian charge density results for this material. The charge density results reveal charge localization of Ti-d electrons and the presence of saddle point at the interface between Se-p and S-p states, which indicate an inter-layer charge transfer in the HS. This observation is further con-firmed by the orbital projected band structure shown in electronic supplementary information (ESI), Fig 1. Here it is clear that the Rashba spin splitting is a combined effect, arising from Se-$p_z$ and S-$p_z$ orbitals located in the valence band (0 to -0.75 eV) while the Ti-$d_{xz}$ and Ti-$d_{yz}$ orbitals are located at the conduction band minima. The origin of the Rashba effect in these HSs is the induced out-of-plane local electric field, which we attribute to removal of the energy degeneracy due to charge polarization of the chalcogen atoms (Te, Se and S). Furthermore, in order to validate the existence of Rashba-type spin splitting, we calculated the spin components of TiSe$_2$/TiS$_2$ HS $s_x$, $s_y$ and $s_z$ at a fixed energy level for the band splitting shown in Fig 3a (right, blue circle). As illustrated in Fig 4(a-c), the spin polarization is more pronounced in the xy-plane: i.e., $s_x$, $s_y$ are more dominant than $s_z$. These spin characteristics show the ideal features of the Rashba effect. The low intensity out-of-plane spin component ( $s_z$) in Fig 4(c) is due to the small in-plane potential gradient responsible for the hexagonal symmetry of the outer contour [43–45].

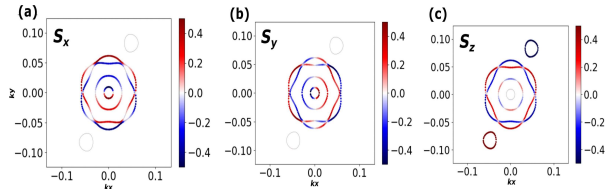

FIG. 4. Spin textures of TiSe$_2$/TiS$_2$ HS a) $s_x$, b) $s_y$ and c) $s_z$ at a constant energy level around Γ point.

### 1. Tuning of Rashba spin-splitting in TiSe$_2$/TiS$_2$

Tuning and enhancing the strength of Rashba spin-splitting via adatom adsorption [46–49] is a well tested method. The aim of introducing an adatom over the surface is to alter the out-of-plane spin polarization. Adatoms with higher electronegativity can create a charge difference at the interface, which creates a local electric field at the surface due to induced interfacial polarization. In this work, we incorporated an F adatom at the surface. Initially, we placed a single F-atom on top of a Se atom or at a hollow site of the surface layer, as shown in Fig 5. Following geometry relaxation of the system with F-adsorbed on top of a Se atom in TiSe$_2$/TiS$_2$ (Fig 5(a-b)), we observe some structural perturbation in the Se-surface layer. Similarly, in the system with F-adsorbed at the hollow site (Fig 5(c-d)), Ti is slightly pulled out-of-plane from its initial position [50]. These structural perturbations in the surface layer, i.e., TiSe$_2$ of the HS, can induce some alterations in the Rashba strength [51, 52]. Energetically, F adsorbed at the hollow site is favored over the alternative adsorption of F at the on-top site [53] which is evident from formation energy results (see ESI Fig 2). The band structures ob-

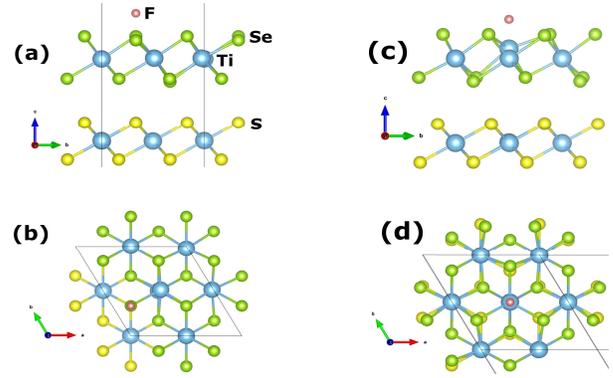

FIG. 5. a) and b) Side and top view of adatom F placed on top of Se; c) and d) Side and top view of F placed at a hollow site of TiSe$_2$/TiS$_2$. Ti appears to be pulled out of position when F is placed at a hollow site of HS.

tained from F-adsorbed TiSe$_2$/TiS$_2$ HS at the top and hollow sites are shown in Fig 6(a,b). When F is located at the top site, attributed to the direct channel from F to Se, the electrons are transferred from F to the surface layer as conduction bands are approaching the valence band region (Fig 6a). However, when F is positioned at the hollow site of the HS surface, it is evident that the valence electrons cross the Fermi level, occupying conduction band states attributed to additional electrons contributed by F. In either adsorption case, metallicity is introduced into the HS by adding an F-adatom. Moreover, the encircled band (brown color in Fig. 6(a,b)) indicates movement of the valence band maxima with the addition of F at different sites. Inspection of the bands also indicate that the F-adatom can tune the Rashba strength. Blue and red circled bands show tuned Rashba strength upon adding the adatom. The Rashba spin splitting obtained in the HS with F at the top site of Se (Fig 6a) are 0.098 ev Å (blue circle) and 0.399 eV Å (red circle) while in the case of F at the hollow site of HS (Fig 6b), it is 0.165 eV Å (blue circle) and 0.759 eV Å (red circle). Although we enjoyed some success in increasing the strength of the Rashba splitting by introducing an adatom over the surface of TiSe$_2$/TiS$_2$ HS, it is important to achieve such effect at the Fermi level for the pruposе of practical applications. We observed that when F is placed at a hollow site of the HS's surface, bands showing Rashba splitting approached the Fermi level. To further push the bands showing Rashba splitting closer to the Fermi level, it could be beneficial to increase the concentration of F-adatoms at hollow sites. To test this hypothesis, we compute the bandstructure with additional F-adatoms. Fig 6(e) shows the band structure obtained following relaxation of the structure of TiSe$_2$/TiS$_2$ HS with *2* F adatoms placed at the hollow site of Se layer of HS. With this approach, we not only attain Rashba spin-splitting of 0.125 eV Å (blown circle) at the Fermi level but also Zeeman spin splitting (blue circle) which can be seen explicitly in ESI Fig 3. Spin textures in ESI

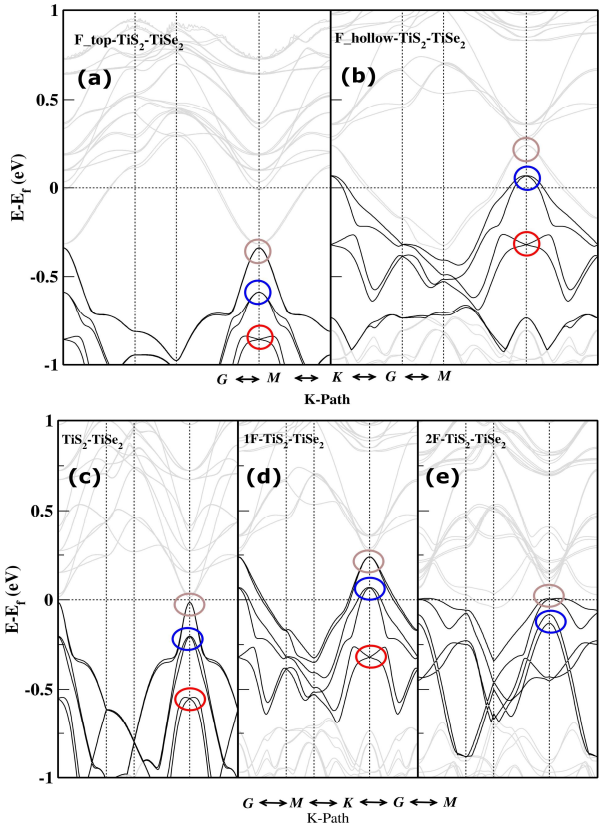

FIG. 6. Electronic band structure with F placed at a) top site of Se layer of HS, b) and d) hollow site of Se layer of HS; c) Band structure of TiSe$_2$/TiS$_2$ HS without adatom; e) 2F adatoms placed at hollow sites of Se layer of HS. Fermi level is set at 0.0 eV. Black color bands show Rashba spin splitting around the Γ point.

Fig 3 represent the Rashba feature at the Fermi level accompanied by Zeeman splitting shown as a pentagon [54]. To address the interaction strength of F and Se/Ti, we performed crystal orbital Hamilton population (COHP) calculations using the LOBSTER code [55]. We calculated the partial COHP and integrated COHP (ICOHP), which provide bond strengths. Negative values of ICOHP indicate strong bonding between the ions. In Fig 7(a), with F placed at the hollow site of HS (over Ti), the ICOHP value at the Fermi energy level (0.0 eV) shows stronger F-Ti bonding than F-Se bonding. Furthermore, in Fig 7(b) when F is located at the top-site (over Se), F-Se bonding is stronger than F-Ti. Moreover, Se-Ti bonding does not seems to vary in either of the cases.

### 2. Existence of Spin-Valley Coupling

Spin and valley degrees of freedom in 2-dimensional TMD systems, are coupled, due to SOC and broken inversion symmetry. To verify such an effect in the TiSe$_2$/TiS$_2$ HS with 2F-adatoms, we performed the Berry curvature calculation [34] along the K'-Γ-K direction. The Berry curvature shown in Fig 8 shows opposite features at the K and K' points, which confirms the coupled spin and valley Hall effects [56, 57].

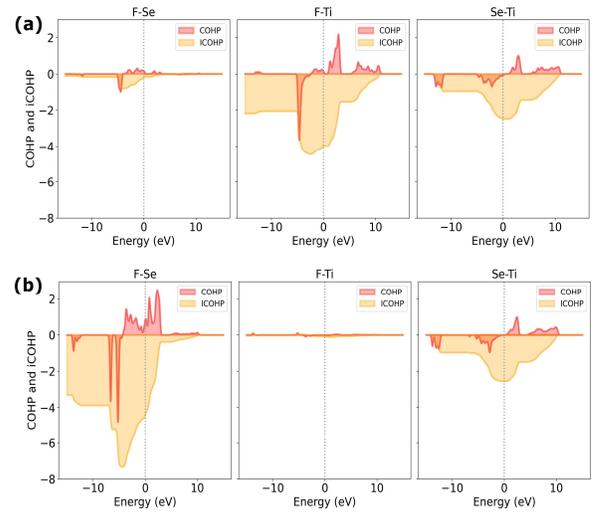

FIG. 7. Partial and intergrated COHP of F-adsorbed at a) hollow site and b) top site of TiSe$_2$/TiS$_2$ HS. Dashed line indicates the Fermi level set at 0.0 eV.

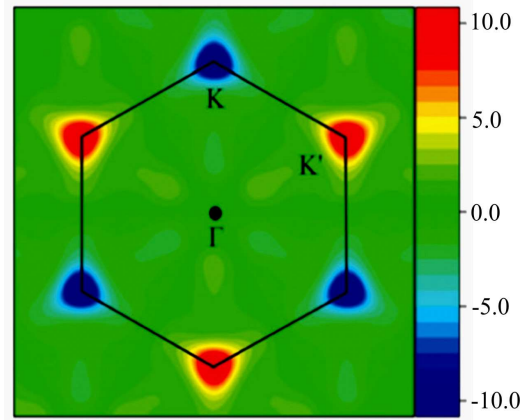

FIG. 8. Berry curvature intensity plot K'-Γ-K high symmetry points showing the presence of spin splitting at the K and K' valleys.

Maximal color gradient with equal and opposite sign indicates the presence of spin up electrons at the K' valley and spin down electrons at the K valley along with in-plane spin electrons at the Γ points. The combination of Rashba spin splitting along the in-plane spin direction and band splitting at the K and K' valleys (i.e., in the out-of-plane spin direction) can be useful in spintronics as well as valleytronics research fields.

## IV. CONCLUSION

In conclusion, we employed DFT simulations accompanied by theoretical analyses to understand the Rashba spin splitting in the Ti-based HSs TiTe$_2$/TiSe$_2$, TiTe$_2$/TiS$_2$ and TiSe$_2$/TiS$_2$. Among the systems, TiSe$_2$/TiS$_2$ showed the most promising Rashba splitting features. In order to further enhance the strength of Rashba splitting, we adsorbed F-ions to the surface of TiSe$_2$/TiS$_2$, either on top or at hollow surface sites. We found that F-atoms adsorbed at hollow sites are more stable. Importantly, F-adsorbed HS showed larger Rashba splitting, and upon increasing the concentration of F-ions, strong Rashba splitting was achieved at Fermi level. With 2F adatoms over the surface, the Rashba effect was observed along with Zeeman splitting which was further confirmed by Berry curvature analysis. Our proposed strategy in this work suggests that one can not only induce Rashba effects, but also enhance its strength. We believe that our current findings might help guide exploration and design of new materials for practical applications in spintronic and valleytronics.

## ACKNOWLEDGMENTS

Author A.B. would like to thank Prof. Mukul Kabir, Department of Physics, Indian Institute of Science Education and Research, Pune, India, for fruitful discussions.

[1] K. Hejazi, Z.-X. Luo, and L. Balents, Noncollinear phases in moiré magnets, Proceedings of the National Academy of Sciences **117**, 10721 (2020).

[2] D. Kiese, F. L. Buessen, C. Hickey, S. Trebst, and M. M. Scherer, Emergence and stability of spin-valley entangled quantum liquids in moiré heterostructures, Phys. Rev. Research **2**, 013370 (2020).

[3] L. Balents, C. R. Dean, D. K. Efetov, and A. F. Young, Superconductivity and strong correlations in moiré flat bands, Nat. Phys. **16**, 725 (2020).

[4] G. Chen, A. L. Sharpe, P. Gallagher, I. T. Rosen, E. J. Fox, L. Jiang, B. Lyu, H. Li, K. Watanabe, T. Taniguchi, J. Jung, Z. Shi, D. Goldhaber-Gordon, Y. Zhang, and F. Wang, Signatures of tunable superconductivity in a trilayer graphene moiré superlattice, Nature **572**, 215 (2019).

[5] Y. Cao, V. Fatemi, S. Fang, K. Watanabe, T. Taniguchi, E. Kaxiras, and P. Jarillo-Herrero, Unconventional superconductivity in magic-angle graphene superlattices, Nature **556**, 43 (2018).

[6] K. Tran, G. Moody, F. Wu, X. Lu, J. Choi, K. Kim, A. Rai, D. A. Sanchez, J. Quan, A. Singh, J. Embley, A. Zepeda, M. Campbell, T. Autry, T. Taniguchi, K. Watanabe, N. Lu, S. K. Banerjee, K. L. Silverman, S. Kim, E. Tutuc, L. Yang, A. H. MacDonald, and X. Li, Evidence for moiré excitons in van der waals heterostructures, Nature **567**, 71 (2019).

[7] N. Zhang, A. Surrente, M. Baranowski, D. K. Maude, P. Gant, A. Castellanos-Gomez, and P. Plochocka, Moiré intralayer excitons in a *MoSe$_2$/MoS$_2$* heterostructure, Nano Letters **18**, 7651 (2018).

[8] K. F. Mak, C. Lee, J. Hone, J. Shan, and T. F. Heinz, Atomically thin *MoS$_2$*: A new direct-gap semiconductor, Phys. Rev. Lett. **105**, 136805 (2010).

[9] Z. Y. Zhu, Y. C. Cheng, and U. Schwingenschlögl, Giant spin-orbit-induced spin splitting in two-dimensional transition-metal dichalcogenide semiconductors, Phys. Rev. B **84**, 153402 (2011).

[10] D. Xiao, G.-B. Liu, W. Feng, X. Xu, and W. Yao, Coupled spin and valley physics in monolayers of *MoS$_2$* and other group-VI dichalcogenides, Phys. Rev. Lett. **108**, 196802 (2012).

[11] A. Rycerz, J. Tworzyd-lo, and C. W. J. Beenakker, Valley filter and valley valve in graphene, Nat. Phys. **3**, 172 (2007).

[12] D. Xiao, W. Yao, and Q. Niu, Valley-contrasting physics in Graphene: Magnetic moment and topological transport, Phys. Rev. Lett. **99**, 236809 (2007).

[13] M.-Y. Li, C.-H. Chen, Y. Shi, and L.-J. Li, Heterostructures based on two-dimensional layered materials and their potential applications, Materials Today **19**, 322 (2016).

[14] G. R. Bhimanapati, Z. Lin, V. Meunier, Y. Jung, J. Cha, S. Das, D. Xiao, Y. Son, M. S. Strano, V. R. Cooper, L. Liang, S. G. Louie, E. Ringe, W. Zhou, S. S. Kim, R. R. Naik, B. G. Sumpter, H. Terrones, F. Xia, Y. Wang, J. Zhu, D. Akinwande, N. Alem, J. A. Schuller, R. E. Schaak, M. Terrones, and J. A. Robinson, Recent advances in two-dimensional materials beyond graphene, ACS Nano **9**, 11509 (2015).

[15] K. S. Novoselov, A. Mishchenko, A. Carvalho, and A. H. C. Neto, 2D materials and van der waals heterostructures, Science **353**, aac9439 (2016).

[16] Y. Liu, N. O. Weiss, X. Duan, H.-C. Cheng, Y. Huang, and X. Duan, Van der waals heterostructures and devices, Nat. Rev. Mater. **1** (2016).

[17] P. Rivera, J. R. Schaibley, A. M. Jones, J. S. Ross, S. Wu, G. Aivazian, P. Klement, K. Seyler, G. Clark, N. J. Ghimire, J. Yan, D. G. Mandrus, W. Yao, and X. Xu, Observation of long-lived interlayer excitons in monolayer *MoSe$_2$ WSe$_2$* heterostructures, Nat. Commun. **6**, 6242 (2015).

[18] M.-K. Lin, T. He, J. A. Hlevyack, P. Chen, S.-K. Mo, M.-Y. Chou, and T.-C. Chiang, Coherent electronic band structure of *TiTe$_2$/TiSe$_2$* moiré bilayer, ACS Nano **15**, 3359 (2021).

[19] D. Pierucci, H. Henck, J. Avila, A. Balan, C. H. Naylor, G. Patriarche, Y. J. Dappe, M. G. Silly, F. Sirotti, A. T. C. Johnson, M. C. Asensio, and A. Ouerghi, Band alignment and minigaps in monolayer *MoS$_2$*-Graphene van der waals heterostructures, Nano Letters **16**, 4054 (2016).

[20] K. S. Novoselov, E. McCann, S. V. Morozov, V. I. Fal'ko, M. I. Katsnelson, U. Zeitler, D. Jiang, F. Schedin, and A. K. Geim, Unconventional quantum hall effect and

berry's phase of $2\pi$ in bilayer graphene, Nat. Phys. **2**, 177 (2006).
[21] G. Dresselhaus, Spin-orbit coupling effects in zinc blende structures, Phys. Rev. **100**, 580 (1955).
[22] Y. A. Bychkov and É. I. Rashba, Properties of a 2d electron gas with lifted spectral degeneracy, JETP lett **39**, 78 (1984).
[23] S. Datta and B. Das, Electronic analog of the electro-optic modulator, Applied Physics Letters **56**, 665 (1990).
[24] G. Bihlmayer, O. Rader, and R. Winkler, Focus on the Rashba effect, New J. Phys. **17**, 050202 (2015).
[25] W. Ju, D. Wang, T. Li, Y. Zhang, Z. Gao, L. Ren, H. Li, and S. Gong, Remarkable Rashba spin splitting induced by an asymmetrical internal electric field in polar III–VI chalcogenides, Phys. Chem. Chem. Phys. **22**, 9148 (2020).
[26] G. Kresse and D. Joubert, From ultrasoft pseudopotentials to the projector augmented-wave method, Phys. Rev. B **59**, 1758 (1999).
[27] P. E. Blöchl, Projector augmented-wave method, Phys. Rev. B **50**, 17953 (1994).
[28] G. Kresse and J. Furthmüller, Efficient iterative schemes for ab initio total-energy calculations using a plane-wave basis set, Phys. Rev. B **54**, 11169 (1996).
[29] C. M. Fang, R. A. de Groot, and C. Haas, Bulk and surface electronic structure of $1t$—$tis_2$ and $1t - tise_2$, Phys. Rev. B **56**, 4455 (1997).
[30] W.-M. Zhao, L. Zhu, Z. Nie, Q.-Y. Li, Q.-W. Wang, L.-G. Dou, J.-G. Hu, L. Xian, S. Meng, and S.-C. Li, Moiré enhanced charge density wave state in twisted 1T-TiTe2/1T-TiSe2 heterostructures, Nat. Mater. **21**, 284 (2022).
[31] S. Grimme, J. Antony, S. Ehrlich, and H. Krieg, A consistent and accurate ab initio parametrization of density functional dispersion correction (dft-d) for the 94 elements h-pu, The Journal of Chemical Physics **132**, 154104 (2010).
[32] U. Herath, P. Tavadze, X. He, E. Bousquet, S. Singh, F. Muñoz, and A. H. Romero, Pyprocar: A python library for electronic structure pre/post-processing, Computer Physics Communications **251**, 107080 (2020).
[33] J. D. Hunter, Matplotlib: A 2d graphics environment, Computing in Science Engineering **9**, 90 (2007).
[34] A. A. Mostofi, J. R. Yates, Y.-S. Lee, I. Souza, D. Vanderbilt, and N. Marzari, wannier90: A tool for obtaining maximally-localised wannier functions, Computer Physics Communications **178**, 685 (2008).
[35] P. Giannozzi, S. Baroni, N. Bonini, M. Calandra, R. Car, C. Cavazzoni, D. Ceresoli, G. L. Chiarotti, M. Cococcioni, I. Dabo, A. D. Corso, S. de Gironcoli, S. Fabris, G. Fratesi, R. Gebauer, U. Gerstmann, C. Gougoussis, A. Kokalj, M. Lazzeri, L. Martin-Samos, N. Marzari, F. Mauri, R. Mazzarello, S. Paolini, A. Pasquarello, L. Paulatto, C. Sbraccia, S. Scandolo, G. Sclauzero, A. P. Seitsonen, A. Smogunov, P. Umari, and R. M. Wentzcovitch, Quantum espresso: a modular and open-source software project for quantum simulations of materials, Journal of Physics: Condensed Matter **21**, 395502 (2009).
[36] Y. A. Bychkov and É. I. Rashba, Properties of a 2D electron gas with lifted spectral degeneracy, Soviet Journal of Experimental and Theoretical Physics Letters **39**, 78 (1984).
[37] T. Etienne, E. Mosconi, and F. De Angelis, Dynamical origin of the Rashba effect in organohalide lead perovskites: A key to suppressed carrier recombination in perovskite solar cells?, The Journal of Physical Chemistry Letters **7**, 1638 (2016).
[38] Q.-F. Yao, J. Cai, W.-Y. Tong, S.-J. Gong, J.-Q. Wang, X. Wan, C.-G. Duan, and J. H. Chu, Manipulation of the large Rashba spin splitting in polar two-dimensional transition-metal dichalcogenides, Phys. Rev. B **95**, 165401 (2017).
[39] W. Zhou, J. Chen, Z. Yang, J. Liu, and F. Ouyang, Geometry and electronic structure of monolayer, bilayer, and multilayer janus WSSe, Phys. Rev. B **99**, 075160 (2019).
[40] J. Chen, K. Wu, H. Ma, W. Hu, and J. Yang, Tunable Rashba spin splitting in janus transition-metal dichalcogenide monolayers via charge doping, RSC Adv. **10**, 6388 (2020).
[41] G. Rehman, S. A. Khan, B. Amin, I. Ahmad, L.-Y. Gan, and M. Maqbool, Intriguing electronic structures and optical properties of two-dimensional van der waals heterostructures of $Zr_2CT_2$ (T = O, F) with $MoSe_2$ and $WSe_2$, J. Mater. Chem. C **6**, 2830 (2018).
[42] G. W. King'ori, C. N. M. Ouma, A. K. Mishra, G. O. Amolo, and N. W. Makau, Two-dimensional graphene–$HfS_2$ van der waals heterostructure as electrode material for alkali-ion batteries, RSC Adv. **10**, 30127 (2020).
[43] L. Xiang, Y. Ke, and Q. Zhang, Tunable giant Rashba-type spin splitting in $PtSe_2/MoSe_2$ heterostructure, Applied Physics Letters **115**, 203501 (2019).
[44] S. LaShell, B. A. McDougall, and E. Jensen, Spin splitting of an Au(111) surface state band observed with angle resolved photoelectron spectroscopy, Phys. Rev. Lett. **77**, 3419 (1996).
[45] Y. M. Koroteev, G. Bihlmayer, J. E. Gayone, E. V. Chulkov, S. Blügel, P. M. Echenique, and P. Hofmann, Strong spin-orbit splitting on Bi surfaces, Phys. Rev. Lett. **93**, 046403 (2004).
[46] E. Rotenberg, J. W. Chung, and S. D. Kevan, Spin-orbit coupling induced surface band splitting in Li/W(110) and Li/Mo(110), Phys. Rev. Lett. **82**, 4066 (1999).
[47] M. Hochstrasser, J. G. Tobin, E. Rotenberg, and S. D. Kevan, Spin-resolved photoemission of surface states of W(110)-(1 1)H, Phys. Rev. Lett. **89**, 216802 (2002).
[48] F. Forster, S. Hüfner, and F. Reinert, Rare gases on noble-metal surfaces: an angle-resolved photoemission study with high energy resolution, The Journal of Physical Chemistry B **108**, 14692 (2004).
[49] L. Moreschini, A. Bendounan, C. R. Ast, F. Reinert, M. Falub, and M. Grioni, Effect of rare-gas adsorption on the spin-orbit split bands of a surface alloy: Xe on Ag(111)-($\sqrt{3} \times \sqrt{3}$)$r30°$−Bi, Phys. Rev. B **77**, 115407 (2008).
[50] R. Friedrich, V. Caciuc, G. Bihlmayer, N. Atodiresei, and S. Blügel, Designing the Rashba spin texture by adsorption of inorganic molecules, New Journal of Physics **19**, 043017 (2017).
[51] R. Friedrich, V. Caciuc, N. Atodiresei, and S. Blügel, Molecular induced skyhook effect for magnetic interlayer softening, Phys. Rev. B **92**, 195407 (2015).
[52] G. Bihlmayer, S. Blügel, and E. V. Chulkov, Enhanced Rashba spin-orbit splitting in BiAg(111) and PbAg(111) surface alloys from first principles, Phys. Rev. B **75**,

195414 (2007).
- [53] C. Zhu and G. Yang, Insights from the adsorption of halide ions on graphene materials, ChemPhysChem **17**, 2482 (2016).
- [54] C. Mera Acosta, A. Fazzio, and G. M. Dalpian, Zeeman-type spin splitting in nonmagnetic three-dimensional compounds, npj quantum mater. **4** (2019).
- [55] R. Nelson, C. Ertural, J. George, V. L. Deringer, G. Hautier, and R. Dronskowski, Lobster: Local orbital projections, atomic charges, and chemical-bonding analysis from projector-augmented-wave-based density-functional theory, Journal of Computational Chemistry **41**, 1931 (2020).
- [56] D. Xiao, G.-B. Liu, W. Feng, X. Xu, and W. Yao, Coupled spin and valley physics in monolayers of $MoS_2$ and other group-VI dichalcogenides, Phys. Rev. Lett. **108**, 196802 (2012).
- [57] Q. Zhang and U. Schwingenschlögl, Rashba effect and enriched spin-valley coupling in $GaX/MX_2$ (M = Mo, W; X = S, Se, Te) heterostructures, Phys. Rev. B **97**, 155415 (2018).

# Electronic Supplementary Information (ESI)
# Coexistence of Rashba Effect and Spin-valley Coupling in TiX$_2$(X= Te, S and Se) based Heterostructures

Amreen Bano*  and  Dan Thomas Major

*banoamreen.7@gmail.com

Department of Chemistry and Institute of Nanotechnology Advanced Materials,
Bar-Ilan University, Ramat Gan 52900, Israel

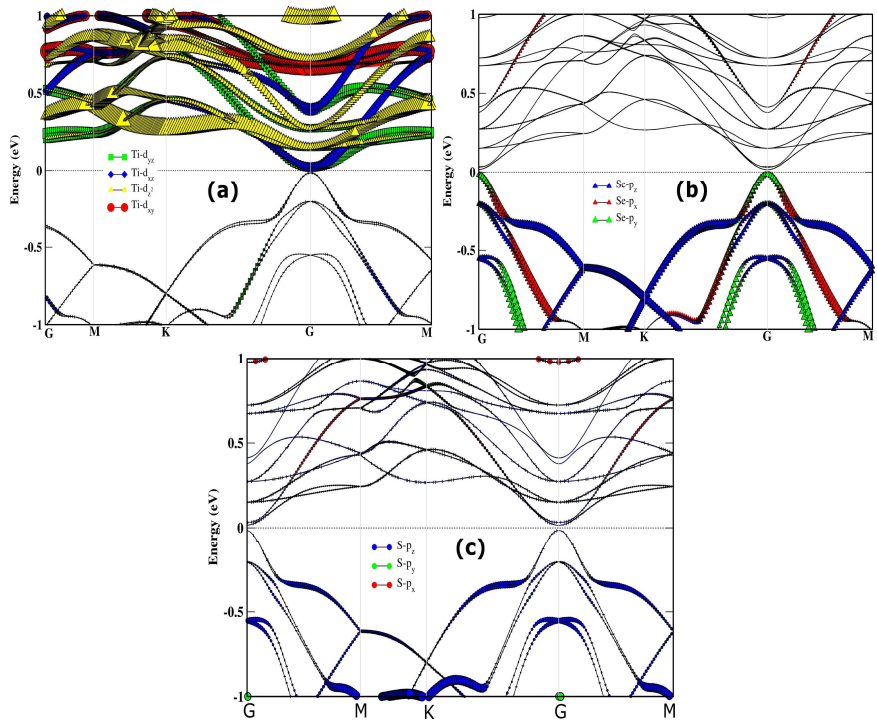

FIG. 1. Orbital projected band structures of a) Ti-d, b) Se-p and c) S-p states of TiSe$_2$/TiS$_2$ heterostructure. Fermi level is set at 0.0 eV.

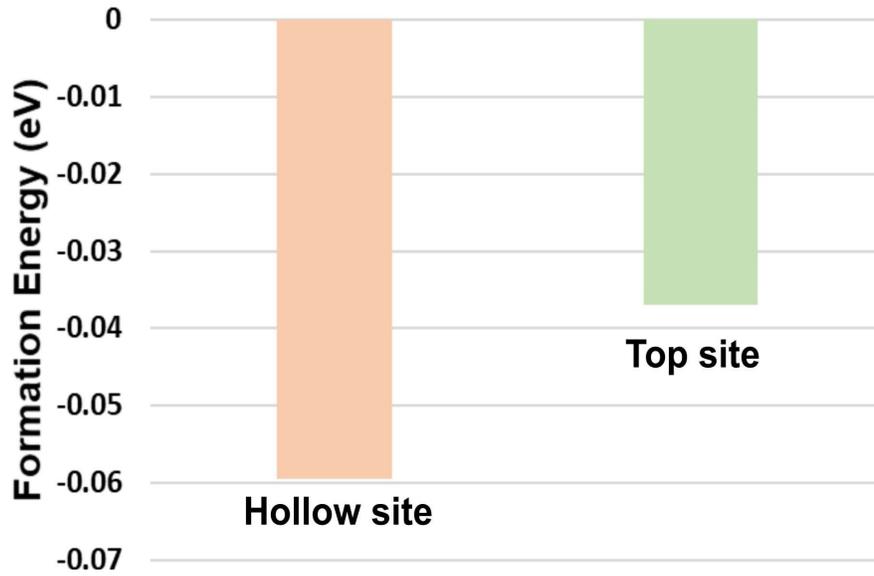

FIG. 2. Formation energy of F-adsorbed at hollow and top sites of TiSe$_2$/TiS$_2$ heterostructure. Lower value for hollow site indicates relatively more stable structure than F at top site of the heterostructure.

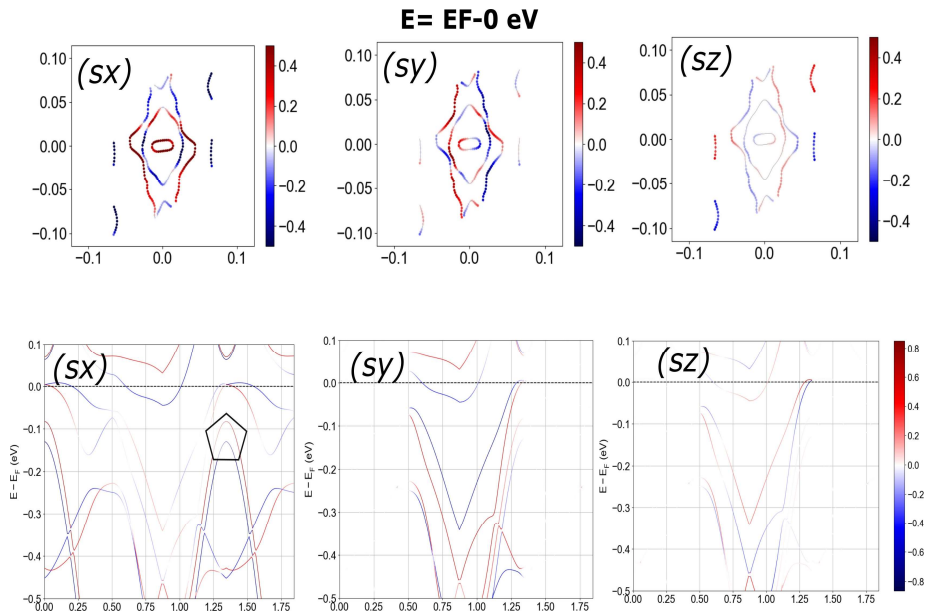

FIG. 3. Top panel shows the spin textures of TiSe$_2$/TiS$_2$ heterostructure with 2F adatoms adsorbed over the surface. It shows the ideal character of Rashba effect having less significant splitting in $s_z$ while considerable spin splitting is observed in $s_x$, $s_y$; Lower panel shows the spin projected band structure of 2F adatoms adsorbed TiSe$_2$/TiS$_2$ heterostructure. Rashba spin splitting (at Fermi level) along with Zeeman splitting (bands in pentagon) coexistence is noticed.